\documentclass[12pt,makeidx,epsfig,floatfig,wrapfig,amsbsy,amsfonts]{article}
\usepackage{times}
\usepackage{amsmath}
\usepackage{graphicx}
\usepackage{epsfig}
\usepackage{wrapfig}

\advance \topmargin by -\headheight
\advance \topmargin by -\headsep
\evensidemargin \oddsidemargin
\begin{document}
\pagestyle{plain}
\begin{titlepage}
\flushright{\today}
\vspace*{0.15cm}
\begin{center}
{\Large\bf
      Search for light pseudoscalar sgoldstino in
  $K^{-}   $ decays  
    }
\vspace*{0.15cm}

\vspace*{0.3cm}
{\bf  
O.G.~Tchikilev, S.A.~Akimenko,  
 G.I.~Britvich, K.V.~Datsko,  A.P.~Filin, 
A.V.~Inyakin,  V.A.~Khmelnikov, A.S.~Konstantinov, V.F.~Konstantinov, 
I.Y.~Korolkov,  V.M.~Leontiev, V.P.~Novikov,
V.F.~Obraztsov,  V.A.~Polyakov, V.I.~Romanovsky, 
  V.I.~Shelikhov, N.E.~Smirnov,   
    V.A.~Uvarov,   O.P.~Yushchenko. }
  
\vskip 0.15cm
{\large\bf $Institute~for~High~Energy~Physics,~Protvino,~Russia$}

\vskip 0.35cm
{\bf 
 V.N.~Bolotov, S.V.~Laptev,   A.R.~Pastsjak, A.Yu.~Polyarush.}
\vskip 0.15cm
{\large\bf $Institute~for~Nuclear~Research~Moscow,~Russia$}
\vskip 0.15cm
\end{center}
\end{titlepage}
\begin{center}
Abstract
\end{center}
 A search for the light pseudoscalar sgoldstino production in the three-body
 $K^{-}$ decay  $K^{-} \rightarrow \pi^{-}\pi^{0}~ P $ has been 
 performed with the ``ISTRA+'' detector exposed to the 25 GeV/c negative 
 secondary beam of the U-70 proton synchrotron. No signal is observed.  
 An upper limit for the branching ratio
 $Br (K^{-} \rightarrow \pi^{-} \pi^{0} P)$, at $90 \%$ confidence level, 
 is determined to be $\sim 9~\cdot~10^{-6}$   in the effective mass 
 $m_{P}$ range  from $0$ to $200$~MeV/c$^2$, excluding the region near 
 $m_{\pi^{\circ}}$ where it degrades to $\sim 3.5~\cdot~10^{-5}$. 
\vskip 0.35cm

\newpage

\section{ Introduction}
\renewcommand{\refname}{References}
\renewcommand{\figurename}{Figure}
\renewcommand{\tablename}{Table}
\begin{wrapfigure}{r}{6.5cm}
\epsfig{file=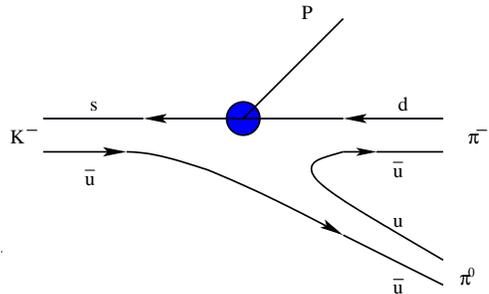,width=6.5cm}
\caption{$K^-$ decay into sgoldstino and pions.}
\end{wrapfigure}

 In  models with spontaneous supersymmetry breaking 
 the superpartners of a Goldstone fermion,
 pseudoscalar $P$ and scalar $S$ goldstinos, should exist. In 
 some versions of gravity-mediated and gauge-mediated theories
  (~for a recent review see \cite{rev}~)  
 one or both of these
 weakly interacting bosons (sgoldstinos) are light enough to
 be observed in kaon decays. Moreover, if sgoldstino interactions with quarks
 conserve parity, as in left-right extensions of MSSM, and $P$ is lighter than
 $S$, so that $m_{S}>(m_{K}-m_{\pi})$ and $m_{P}<(m_{K}-2m_{\pi})$, sgoldstinos 
 can be observed in the decay $K \rightarrow \pi \pi P$(see Fig.1), rather than 
 in  the much better constrained $K \rightarrow \pi S$.
 The phenomenology of light sgoldstinos in this scenario is considered in
 detail in \cite{ref1}. 
 Under the
 assumption  that sgoldstino interactions with quarks and
 gluons violate quark flavor and conserve parity, low energy interactions
 of pseudoscalar sgoldstino $P$ with quarks are described by the Lagrangian:
\begin{equation}
 L = -P \cdot (h^D_{ij} \cdot \overline{d}_i i \gamma^5 d_j +
 h^U_{ij} \cdot \overline{u}_i i \gamma^5 u_j)~,
\end{equation}
 where
\begin{center}
$  d_i = (d,s,b)~, ~~~~~~ u_i = (u,c,t)~,$
\end{center}
  and with coupling constants $h_{ij}$ proportional to the left-right soft
 terms in the mass matrix  of squarks:
\begin{equation}
   h^{D}_{ij} = \frac{\mbox{\~{m}}^{(LR)2}_{D,ij}}{\sqrt{2} F}~, ~~~~~~~~
   h^{U}_{ij} = \frac{\mbox{\~{m}}^{(LR)2}_{U,ij}}{\sqrt{2} F}~,
\end{equation}
 where the scale of supersymmetry breaking is denoted as $\sqrt{F}$. \\ 
  The $90\%$ confidence level~(CL) 
  constraints on the flavor-violating coupling of sgoldstinos
  to quarks   evaluated using the $K^{\circ}_L - K^{\circ}_S $ 
mass difference  and $CP$ violating parameter $\epsilon$ in the neutral kaon 
system are:
 $|h^{D}_{12}| \leq 7~\cdot~10^{-8}$; $
 |\mbox{Re}(h^{D}_{12}) \mbox{Im}(h^{D}_{12})| < 1.5~\cdot~10^{-17} $ .
 It has been shown  \cite{ref1}
   that, depending on the phase of sgoldstino-quark coupling, these
constraints result in the following $90\%$ CL upper 
limits on the branching ratio: 
$Br(K^{-} \rightarrow~\pi^{-}\pi^{0}P) 
 \leq1.5~\cdot~10^{-6}~$---$~4~\cdot~10^{-4}$,  where the less strong limit
  corresponds to the  case of pure real or pure imaginary $h^{D}_{12}$ . 
  A search for $P$ in 
 charged kaon decays is of particular interest when 
 $\mbox{Re}(h^{D}_{12}) \sim 0$, and 
 when related branching ratio for the decay
 $K_{L} \rightarrow~\pi^{-}\pi^{+}P$ is
 small.  

 Light sgoldstinos decay into two photons or into a pair of charged leptons. 
 The two photon decay dominates almost everywhere in the parameter space. 
 Depending on the parameter $g_{\gamma}=
 \frac{1}{2\sqrt{2}}\frac{M_{\gamma \gamma}}{F}$, where $M_{\gamma \gamma}$
 is the photino mass, sgoldstinos will have a very different lifetime.
 In the present search we assume that the 
 sgoldstino is sufficiently long lived to
 decay outside the detector, i.e. the sgoldstino is ``invisible''.
  The existing $90\%$ CL upper limit on the 
  branching ratio $Br (K^{-}\rightarrow \pi^{-} \pi^{0} P)$
is  $ 4~\cdot~10^{-5}$  \cite{ref6}, whereas the 
$90\%$~CL upper limits for the
scalar sgoldstino $S$   estimated from the
studies of the $K^+ \rightarrow \pi^+ \nu \overline{\nu}$~ vary between 
 $0.4~\cdot~10^{-10}$ 
  and $1.0~\cdot~10^{-10}$ in the mass interval from 0 to 110~MeV/c$^2$,
 see Fig.~3 in  \cite{add2}).
  The recent upper
 limit for the decay
  $K^+ \rightarrow \pi^+ X^{\circ}$, where $X^{\circ}$ is a
 neutral weakly interacting massless particle, is 
 $0.73 \cdot~10^{-10}$($90\%$~CL)\cite{anis}.

  The aim of our present   study  
 is to search for invisible pseudoscalar sgoldstinos in the 
  $K^{-}$ decays. The experimental setup and event selection are described
 in section~2, the results of the analysis are presented in section~3,
 the systematic uncertainties are discussed in section~4
 and the conclusions are given in the last section.

\section{ Experimental setup and event selection}

\begin{figure}
\epsfig{file=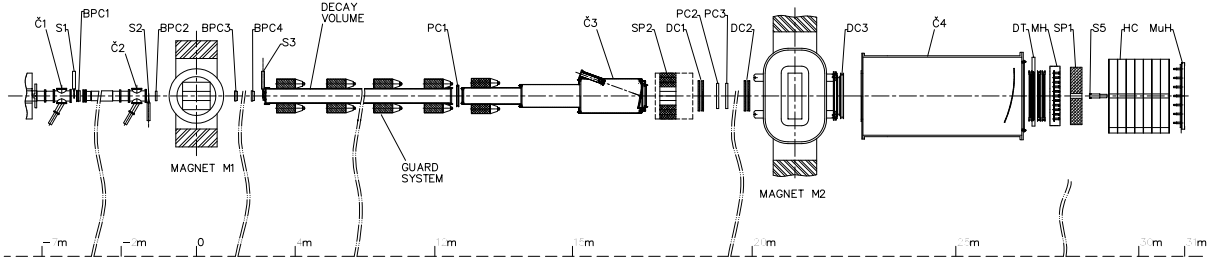,height=16cm,angle=90}
\caption{The elevation view of the ISTRA+ setup. M$_1$ and M$_2$ are magnets,
\v{C}$_i$ --- Cerenkov counters, BPC$_i$ --- beam proportional chambers,
 PC$_i$ --- spectrometer proportional chambers, SP$_i$ --- lead glass
 calorimeters, DC$_i$ --- drift chambers, DT$_i$ --- drift tubes,
 HC --- a hadron calorimeter, S$_i$ --- trigger scintillation counters,
 MH --- a scintillating hodoscope and MuH --- a scintillating
 muon hodoscope.}
\end{figure}

The experiment is performed at the IHEP 70 GeV proton synchrotron U-70.
The    ISTRA+ spectrometer has been described in some detail in our
recent papers on $K_{e3}$  \cite{ref2}, $K_{\mu 3}$ \cite{ref3} 
  and $\pi^-\pi^{\circ}\pi^{\circ}$ decays \cite{ref4}.
Here we recall briefly  the characteristics relevant to our analysis. 
 The  ISTRA+ setup is located in a negative unseparated 
 secondary beam line 4A of U-70. The beam momentum  is $\sim 25$ GeV/c with 
$\Delta p/p \sim 1.5 \%$. The admixture of $K^{-}$ in the beam is $\sim 3 \%$,
 the beam intensity is $\sim 3~\cdot~10^{6}$ per 1.9 sec U-70 spill.
 A schematic view of the ISTRA+ setup is shown
 in Fig.~2. The  beam particles are deflected by the magnet M$_1$
 and are measured by four proportional chambers BPC$_1$---BPC$_4$ with 1~mm
 wire spacing, the kaon identification is done by three threshold Cerenkov
 counters \v{C}$_0$---\v{C}$_2$. The 9~meter long vacuum decay
 volume is surrounded by eight lead glass rings used to veto low energy
 photons. The  72-cell lead-glass calorimeter SP$_2$ plays the same role.
 The decay products are deflected in the magnet M$_2$ with 1~Tm field integral
  and are measured with 2~mm step proportional chambers PC$_1$---PC$_3$, with
 1~cm cell drift chambers DC$_1$---DC$_3$ and, finally, with 2~cm diameter
 drift tubes DT$_1$---DT$_4$. The wide aperture threshold Cerenkov counters
 \v{C}$_3$~,\v{C}$_4$~, filled with He,  serve to trigger electrons
 and are not used in the present measurement. SP$_1$ is a
 576-cell lead-glass calorimeter, followed by HC, a scintillator-iron
 sampling hadron calorimeter. MH is a 11x11 cell scintillating hodoscope,
 used to  improve the time
 resolution of the tracking system, MuH is a 7x7 cell muon hodoscope.
 
  The trigger is provided  by scintillation counters S$_1$---S$_5$,
  beam Cerenkov counters and by the analog sum of amplitudes from
  last dynodes of the SP$_1$ : 
   T=S$_1 \cdot \mbox{S}_2 \cdot \mbox{S}_3 \cdot \overline{\mbox{S}}_4\cdot 
   \mbox{\v{C}}_1
   \cdot \overline{\mbox{\v{C}}}_2 \cdot 
   \overline{\mbox{\v{C}}}_3 \cdot
   \overline{\mbox{S}}_5 \cdot \Sigma(\mbox{SP}_1)$,
   here S$_4$ is a scintillation counter with a hole
   to suppress the beam halo, S$_5$ is a counter downstream of the setup at the
   beam focus, $\Sigma$(SP$_1)$   requires that the analog sum to be
   larger than the MIP signal.
  
 During  the first run in March-April 2001, 363 million of
 trigger events were recorded.
 During the second physics run in November-December 2001 350 million trigger
 events were collected with  higher beam intensity and  stronger
 trigger requirements. This information
 is complemented by about 300~M  Monte Carlo~(MC) events generated using 
 Geant3 \cite{ref5}
 for the dominant $K^-$ decay modes. Signal efficiency for possible sgoldstino
  production has been estimated using 1.5 M generated
  events; in this sample  the sgoldstino mass
   $m_{P}$  is varied from 0 to 200~MeV/c$^2$ in steps of 10~MeV/c$^2$.
  These simulated signal events were weighted using the matrix element given
  in \cite{ref1}.
    
\begin{wrapfigure}{r}{8.1cm}
\center{
\epsfig{file=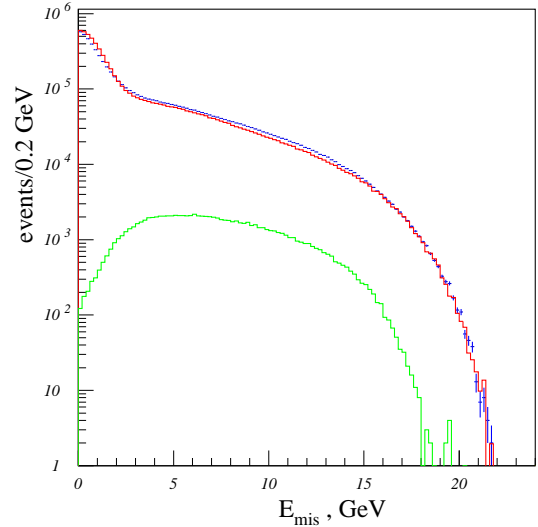,width=8.1cm}
\caption{Missing energy spectra, second run data. Points show
 real data, upper histogram --- background MC events, lower
 histogram --- signal MC events for $m_{P}=90$~MeV/c$^2$.}
 }
\end{wrapfigure}
 Some information on the data
 processing and reconstruction procedures is given in \cite{ref2,ref3,ref4},
 here we briefly mention the details relevant for 
$\pi^{-}\pi^{\circ}$~+~missing energy  event selection. 

 The muon identification (see \cite{ref3}) is based on the information from the
  SP$_1$  and the HC. The electron identification (see \cite{ref2})
 is done using the $E/p$ ratio ---
  the energy  of the shower associated with a charged track and 
  the charged track momentum.
   
  A set of cuts is developed  to suppress 
 backgrounds to possible sgoldstino production:

 0) Events  with one reconstructed charged track and with two
  showers  in the  calorimeter SP$_1$ are selected.
 We require  the effective mass $m(\gamma\gamma)$ to be within
 $\pm 50$~MeV/c$^2$ from $m_{\pi^{\circ}}$. 
  Events with a vertex inside  $ 400 < z < 1650$~cm are selected.

1)  ``Soft'' charged pion identification is applied: tracks identified as 
 electron or muon (as described in \cite{ref2,ref3}) are rejected.
 
2)   Events with  missing energy
 $E_{mis}=E_{beam} - E_{\pi^{-}} - E_{\pi^{\circ}}$ above  3~GeV are selected. 
 This cut  serves to reduce the $K_{\pi2}$ contamination.  
 The missing energy spectra  for the second run data, MC background and
 MC signal with $m_{P}=$90~MeV/c$^2$ are compared in Fig.~3.

3)  To reduce the  background caused by
 secondary charged particles in the photon sample, 
 the MH information is used.
 The distribution of the distance $\Delta r$  between an SP$_1$ shower and the nearest MH hit 
 in the  plane transverse to the beam is shown in
 Fig.~4. An  event is selected if at least one shower has 
  $\Delta r$ greater than 10~cm. 

4) The events where one of the photons is suspected to be irradiated by 
   the charged particle 
   in a detector material upstream/inside  M2-magnet are rejected. 
 Such  photons have nearly the same $x$ coordinate (in the direction
 of the magnetic field) as the charged track on the SP$_1$, i.e.
 an event is rejected if for one or both  showers have the difference
 $\Delta x = | x_{ch} - x_{\gamma} | < 7$~cm .

5)  Events which pass  $K_{e3}$ 2C-fit~\cite{ref2}
are removed.

6)  An additional cut on  the 
 charged pion momentum $p^*$ in the kaon
 rest frame $p^*(\pi^-) < 180$~MeV/c is applied to suppress the tails from
 the $K_{\pi2}$ decay.

\begin{wrapfigure}{r}{7.5cm}
\epsfig{file=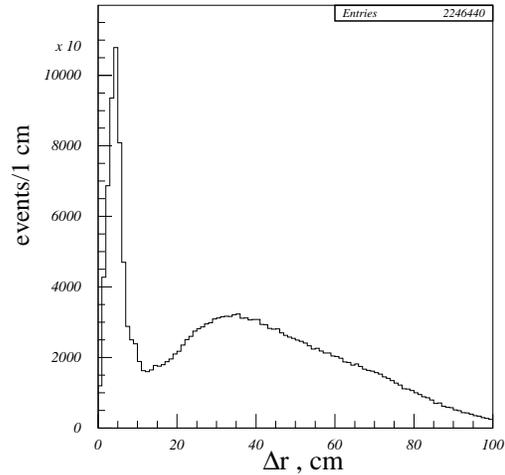,width=7.5cm}
\caption{The distance $\Delta r$ between the SP$_1$ shower and the nearest
 MH hit, second run data.}
\end{wrapfigure}
 
7)  The  pion 
 identification, mainly based on the HC information is required .
 The efficiency and muon suppression of this selection has been determined 
 using $K_{\mu 2}
 $ and
 $K_{\pi 2}$ decays. The efficiency  is found to be  70\% and 80\% 
 for the first and second run data respectively.
 The remaining muon admixture is equal to 3\% and 2\%.

8)    The events with missing momentum pointing to the
    SP$_1$ working aperture are selected in order to suppress
 $\pi^-\pi^{\circ}\gamma$ background.

9) The  ``Veto'' cut uses information from the Guard System (GS) and the guard 
electromagnetic calorimeter (SP$_2$): absence of  signals
above noise threshold is required.


The  data reduction information for the second run 
is given in Table~1  and is compared  with the
 MC-background   and  with
  the MC-signal  for the sgoldstino mass of $m_P=90$~MeV/c$^2$. 
 The cut suppression factors $w_{i-1}/w_i$ for the MC
 data are calculated using the corresponding matrix elements. 
   
The influence of the last  cuts on the missing mass squared spectra
 $(P_{K}-P_{\pi^{-}}-P_{\pi^{\circ}})^{2}$
 is shown in Fig.~5.
 The left wide bump in Fig.~5 is due to 
 $K_{\mu3}$  decays, the shift to the negative missing mass squared 
  is caused by the use of the pion mass
 in its calculations. The second peak is caused by 
 $\pi^{-}\pi^{\circ}(\pi^{\circ})$ decays
 with photons from the second $\pi^{\circ}$ escaping  detection both 
 in the SP$_{1}$
 and in the ``Veto'' system. 
\begin{table}[tbhp]
\caption{ Event reduction statistics for the 2$^{nd}$ run,
 the data, the background MC and the signal MC with $m_P=90$~MeV/c$^2$. } 
\renewcommand{\arraystretch}{1.2}
\begin{center}
\begin{tabular}{|l|c|c|c|c|c|c|}
\hline
 Cut  &  data& $N_{i-1}/N_{i}$ &  BG MC &
 $w_{i-1}/w_{i}$& signal MC& $w_{i-1}/w_{i}$  \\  
\hline
0) 1 $\pi^{-}$, $m(\gamma\gamma)$ near $m(\pi^{\circ}$) 
  & 9943046 &  &5512890 & & 98289 &\\
\hline
1)  no (e, $\mu$)  & 7771606 & 1.28 & 4545059 & 1.19 & 93470 & 1.05\\
\hline
2) E$_{mis}>3.0$~GeV  &  1123220 & 6.92 &588735 & 7.78 & 82602 & 1.17\\
\hline
3) MH filter   & 939052 & 1.20 & 516922 & 1.19 & 74744 & 1.09\\
\hline 
4) conv. gammas  & 722622 & 1.30 & 426286 & 1.22 & 56513 & 1.25    \\
\hline
5) no $K_{e3}$  fit &  458338 & 1.58 &201580 & 2.27 & 35906 & 1.69\\
\hline
6) $p^*(\pi^-) <180$~MeV/c & 326935 & 1.40 & 134706 & 1.28 & 35698 & 1.01\\
\hline
7) $\pi^-$ identification & 122804 & 2.66 & 68380 & 2.06 & 33401 & 1.06\\
\hline
8) $10<r<60$~cm & 108992 & 1.13 & 60698 & 1.12 & 31431 & 1.06 \\
\hline
9) Veto & 31451 & 3.47  & 18674 & 3.58 & 31104 & 1.01\\
\hline
\end{tabular}
\renewcommand{\arraystretch}{1.0}
\end{center}
\end{table}
\section{Analysis and results}

As a result of the previous cuts, especially the last ``Veto'' cut,
the $\pi ^{\circ}$-signal practically disappears 
from the $\gamma \gamma$ mass spectrum for certain missing mass ($m_P$)
intervals. To illustrate that,  the $m(\gamma\gamma)$ spectrum after 
cut (2) is shown in Fig.~6~(left half).
The right half of Fig.~6 shows the $\gamma \gamma$ mass spectra 
 after the 
 ``Veto'' cut for several $m_P$ intervals. It is clearly seen that
 the $\pi^{\circ}$ signal survives for the region of the
 $\pi^-\pi^{\circ}(\pi^{\circ})$ decay only.  
\begin{figure}
\center{
\epsfig{file=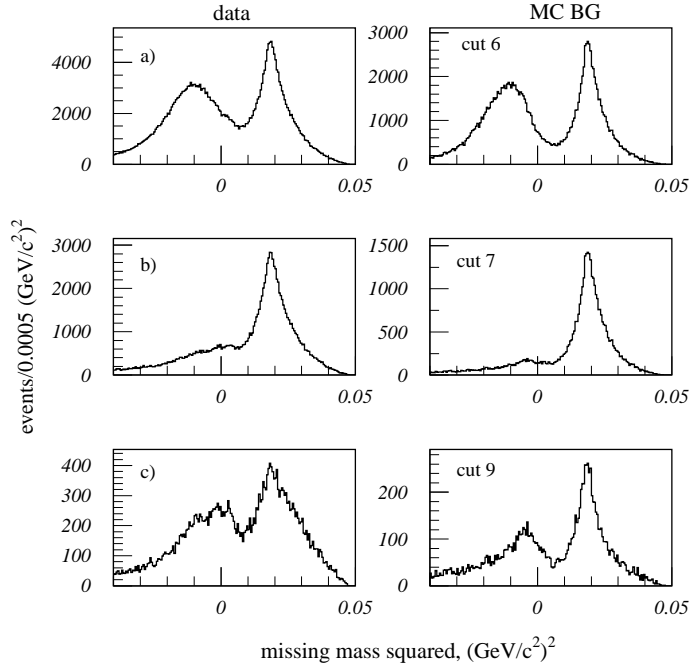,height=9.9cm}
\caption{Missing mass squared distributions after cuts 6~(a) , 7~(b)
  and 9~(c), 
 left column --- second run data, right column --- MC for dominant $K^-$
 decay modes.
 The  bin size is equal to 0.0005 (GeV/c$^{2})^2$. }
}
\end{figure}
 This situation allows us to perform an effective 
background subtraction procedure: 
\begin{figure}
\begin{tabular}{cc}
\epsfig{file=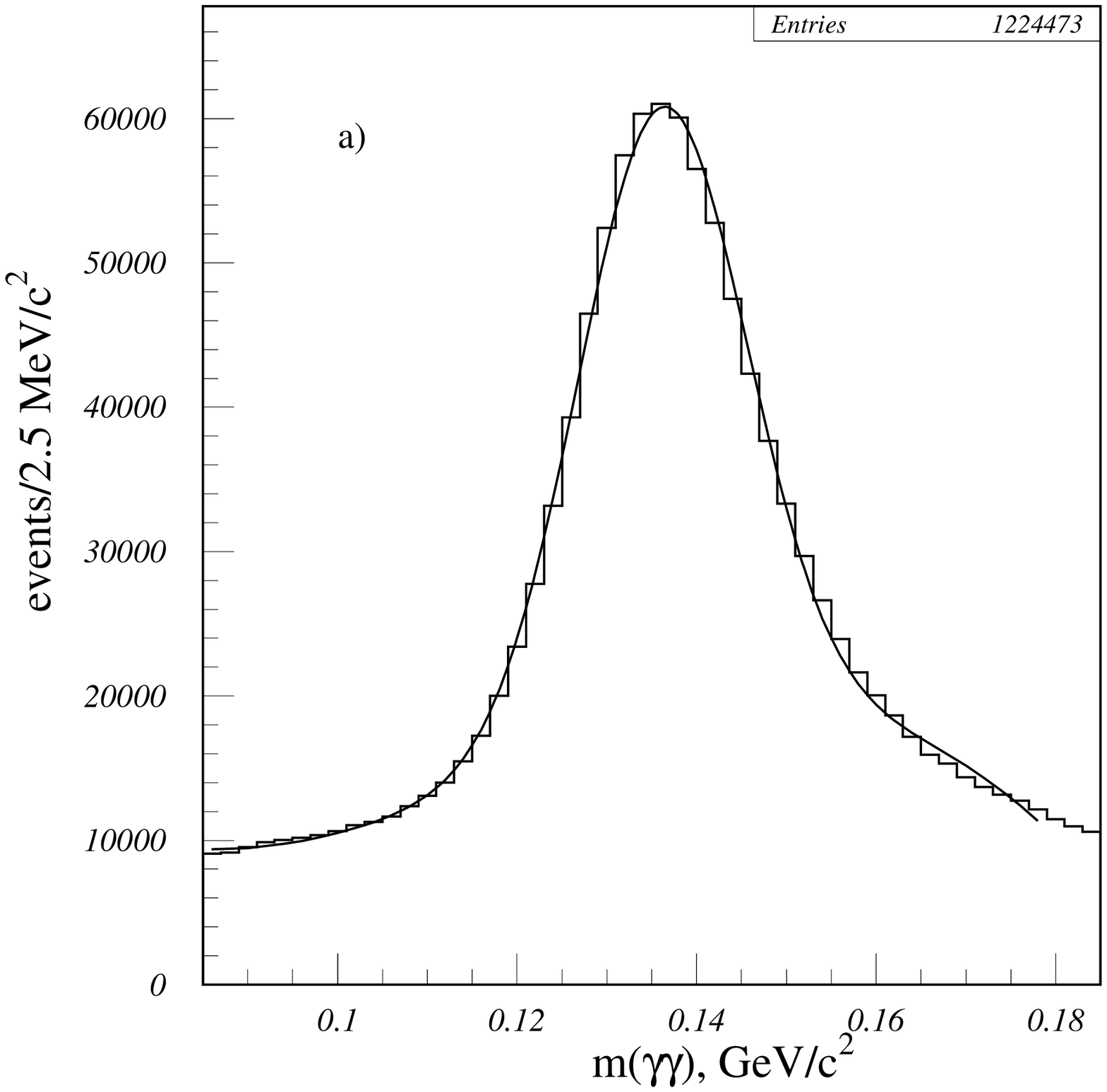,width=8.5cm} &
\epsfig{file=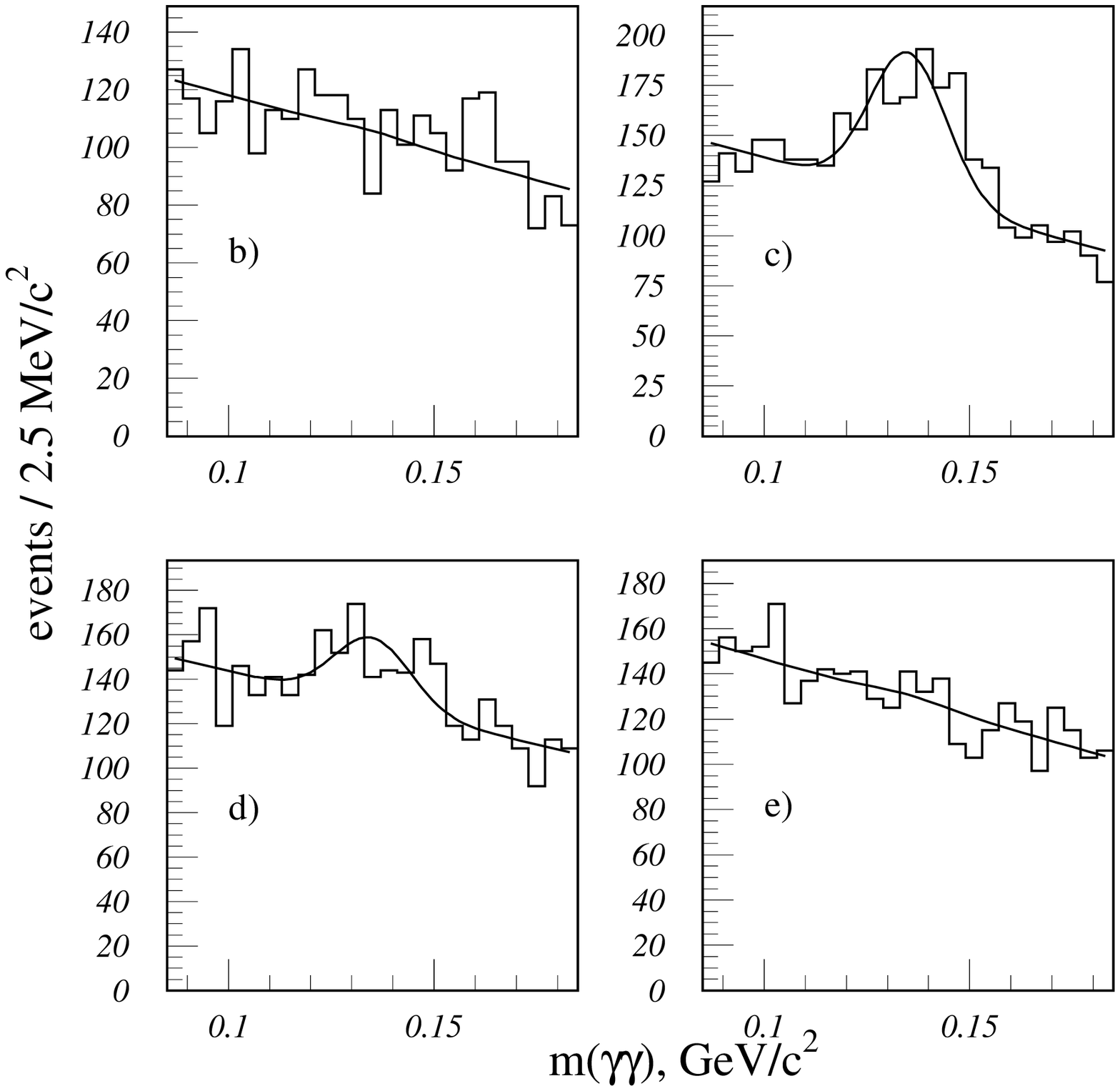,width=8.5cm}
\\
\end{tabular}
\caption{a) $m(\gamma\gamma)$ spectrum for events with $E_{mis} > 3.0$~GeV
  with the result of the fit by the sum of the Gaussian
 and a third degree polynomial; 
 $m(\gamma\gamma)$ spectra after the ``Veto'' cut
  for the missing mass $m_P$ intervals   120-130(b),
   130-140(c), 140-150(d) and
  150-160~MeV/c$^2$(e). }
\end{figure}
 the $m(\gamma\gamma)$ spectra for 10 MeV/c$^2$ $m_P$  intervals, starting from
 $m_P=0$,  are fitted
by the sum of a Gaussian (with the fixed width of 9.3~MeV/c$^2$ and the
 fixed mass of 135~MeV/c$^2$) and a quadratic (or linear) 
polynomial. Such fits are  shown in Fig.~6~(right half). Fig.7 shows 
the obtained numbers of  $\pi^{\circ}$ ($N_{\pi^{0}}$) plotted in  
the corresponding missing  mass~($m_P$) bins. A possible constraint  
 $N_{\pi^{0}}>0$ is not imposed during the fit, thus  
$N_{\pi^{0}}$ is negative for some bins. This was done to avoid a 
specialized treatment of
asymmetric errors during  later analysis.  
The  sgoldstino signal would look like a peak in this 
``filtered'' missing mass spectrum of Fig.7, similar to that of a well seen
$\pi^{0}$ peak (at 135 MeV/c$^2$) from the 
$K^{-} \rightarrow  \pi^-\pi^{\circ}(\pi^{\circ})$  
events which pass the ``Veto'' cut.

No significant signal is observed. To obtain the upper limits for 
the number of sgoldstino 
events at different $m_P$ ($N_{sig}$), the  
"filtered" missing mass spectrum is fitted by the sum of the  signal
Gaussian (with a fixed width of $\sigma =11.1$~MeV/c$^2$, as determined from 
the signal MC) and a  background. 
The background is described by the sum of two components: 
 a Gaussian for the $\pi^-\pi^{\circ}(\pi^{\circ})$ peak
 plus the $\mbox{constant~term}$.
The integral of the signal+background function  over a particular bin (-i-) 
enters into the $\chi ^{2}$ function 
 which is minimized by the ``MINUIT'' program \cite{min}.
 Fig.7 shows an
example of such a fit with   $m_P=185$~MeV/c$^2$.
  To be consistent, we  do not impose the constraint $N_{sig}>0$.
 We checked that the constrained fit
gives comparable or smaller error estimates~(corresponding to lower upper
limits), i.e. the procedure used is more
``conservative''.
One-sided upper limits for the number of signal events at the
90$\%$ confidence level are calculated as
\begin{equation}
 N_{UL} = max(N_{sig},0) + 1.28 \cdot \sigma ~~~,
\end{equation}
  where  $\sigma$ is an  error estimate for $N_{sig}$. 
  The  upper limit for the branching ratio ($UL$) is calculated as
  \begin{equation}
 UL =  
  \frac{N_{UL} \cdot 0.2116  \cdot \varepsilon(K_{\pi 2})}
 { N(K_{\pi 2}) \cdot \varepsilon}
\end{equation}
 where $0.2116$ is  the branching ratio $Br(K_{\pi 2})$,
  and
 $N(K_{\pi 2})$ is the number of reconstructed 
 $K^{-} \rightarrow \pi^-\pi^{\circ}$ decays
 found to be $\sim 1.5$~M events for the first and $\sim 4.5$~M events 
 for the second run.
 The values $\varepsilon$ and $\varepsilon(K_{\pi 2})$ are respective
 efficiencies for $K^{-} \rightarrow \pi^-\pi^{\circ}~P$ and $K_{\pi 2}$ decays,
 which include both the reconstruction efficiency and geometrical acceptance.
 To avoid the systematics associated with the ``Veto'' cut (cut~9), we have applied it
 to the
 selected $K_{\pi 2}$ events. The measured efficiency of this cut is $90~\%$ and is
 explained by random signals in the Veto system due 
 to the presence of the beam halo and $\mbox{event~overlaps}$.
\begin{wrapfigure}{r}{9cm}
\begin{center}
\epsfig{file=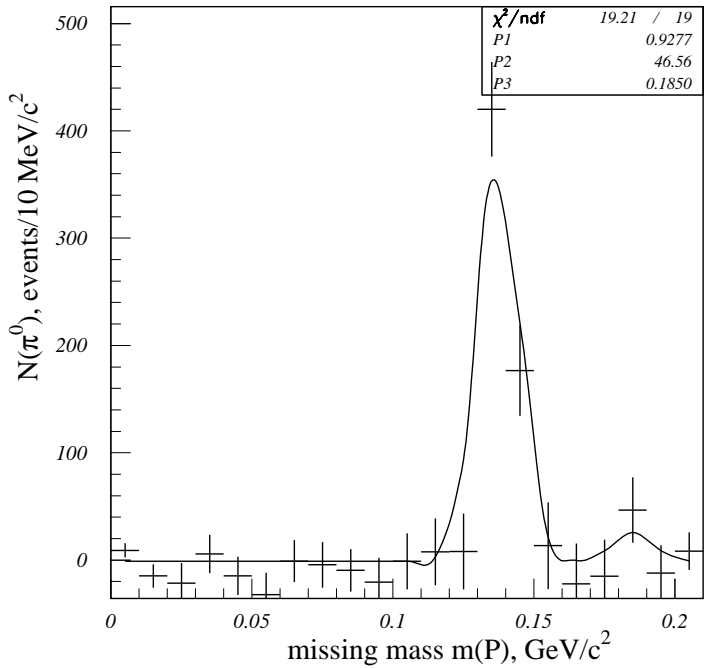,width=10.5cm}
\caption{ The  number  of $\pi^{\circ}$  from the fit of 
 the  $m(\gamma \gamma$) mass spectra plotted in the corresponding
  missing mass $m_P$ bins. The curve is
 the result of the fit with an assumed sgoldstino
  signal at  $m_P=185$~MeV/c$^2$ .}
\end{center}
\end{wrapfigure}
 Since the Veto inefficiency has the same influence on both 
 the signal and  the normalization sample, it cancels out in (4).
 The signal efficiency $\varepsilon$ rises monotonically from
 $\sim 3.3\%$ to $\sim 11.3\%$ in the mass interval from 0 to 170~MeV/c$^2$
 and then drops slightly to the value $\sim 10.3\%$ at 200~MeV/c$^2$.
  This behavior is practically identical for both runs.
 The $K_{\pi 2}$ efficiency
  $\varepsilon(K_{\pi 2})$ is equal to $15.8~\%$ for
 the first run and $22.4~\%$ for the second run. 
 The weighted average ratio 
  $\varepsilon/\varepsilon(K_{\pi 2})$ 
  with the weights proportional to
  the run statistics is used  for the combined data sample.
 
 The final results obtained with the combined statistics of two runs are given
 in Fig.~8.  The  left part of this figure shows a comparison of our results 
   with that published by the E787 collaboration~\cite{ref6}. 

  The obtained upper limits can be transformed into the limits for the value
 of the modulus $| h^D_{12}|$ (see equations~1,2). The corresponding limits 
 are compared in Fig.~8~(right part) with the theoretical limit evaluated using
 $K_L - K_S$ mass difference. A generalization of the formula B9 in the 
  Appendices  B of the paper~\cite{ref1} for the non-zero $m_P$ \cite{Gorbi} 
 was used  when extracting the  $| h^D_{12}|$.
 
  The systematics in the upper limits
 is discussed in the following section.

\section {Systematics }
 The comparison of our data with the background MC shows some discrepancies:
 although the total suppression factors of the selection cuts for 
 the data and MC are equal within a few percent, a significant difference of 
 25-30 $\%$ is observed for cuts 6,7(see Table.1). Some differences
 seen in Figs 3 and~5 are explained by the 
 non-Gaussian tails (absent in the MC data)
 of the beam momentum distribution and higher noise levels in the real data.
 
 To estimate the systematics related to the observed differences,
  we have calculated from our data
  the well known branching ratio 
 $Br(K^{-} \rightarrow \pi^-\pi^{\circ} \pi^{\circ})$, using the events of this
 decay with one $\pi^{\circ}$ missed, i.e. events with the topology
  $K^- \rightarrow \pi^{-} \pi^{\circ} (\pi^{\circ})$.
   In this test the missing $\pi^{\circ}$  plays 
  the role of sgoldstino.
  
   For the test, we have repeated the every step of our 
  analysis, i.e. event selection, which includes: cuts 0-8, filtering of the 
  missing mass spectrum, fitting and normalizing to the
  $K^{-} \rightarrow \pi^-\pi^{\circ}$ decay. The only cut which was removed is
  the ``Veto'' cut (cut 9) in order not to suppress the escaping $\pi^{\circ}$. The 
   branching ratio $Br(K^{-} \rightarrow \pi^-\pi^{\circ} \pi^{\circ})$ has
   been measured to be $ 1.64 \pm 0.03\%$(stat.)  compared with the PDG
   value $1.73 \pm 0.04~\%$~\cite{PDG}, i.e. the observed systematics is 
   $\sim 5 \%$. It has been taken into account by the multiplication
   of the UL in (4) by the factor of 1.05 for all data points. 
   
 
  This procedure ensures the  absence of the systematics in the upper limit for
  sgoldstino mass $m_P \simeq \mbox{m}_{\pi ^{\circ}}$. To estimate the residual systematics for
  other mass bins the upper limits were recalculated for:
  
a) different cut levels in the cuts 2 --- 9;

b) with a different background parametrization for $m(\gamma\gamma)$ (Fig.~6) and
  $N_{\pi^{\circ}}$ (Fig.~7) spectra;
  
c) variation of the theoretical parameters~\cite{ref1}
 used in the signal Monte-Carlo program.   

 For each new set of cuts and for each mass point $m_P$
 the upper limits were calculated exactly in the same
 way as for the main set of cuts. The observed variation of the results gives the 
 estimate of the remaining systematics. For example, for $m_P = 95~$MeV/c$^2$ we
 got $17\%$, $11\%$ and $4\%$ variation for the contributions
  a), b) and c) respectively.
 
  This systematics for each point
  was taken into account by the corresponding increase
  (by the  summary multiplicative factor) of the upper
  limits ( $21\%$ increase for $m_P = 95$~MeV/c$^2$
  equal to sum of the contributions a), b) and c)).  For other data points the
  systematics varies from 16 to 25~\%.  
  
\begin{figure}
\begin{tabular}{cc}
\epsfig{file=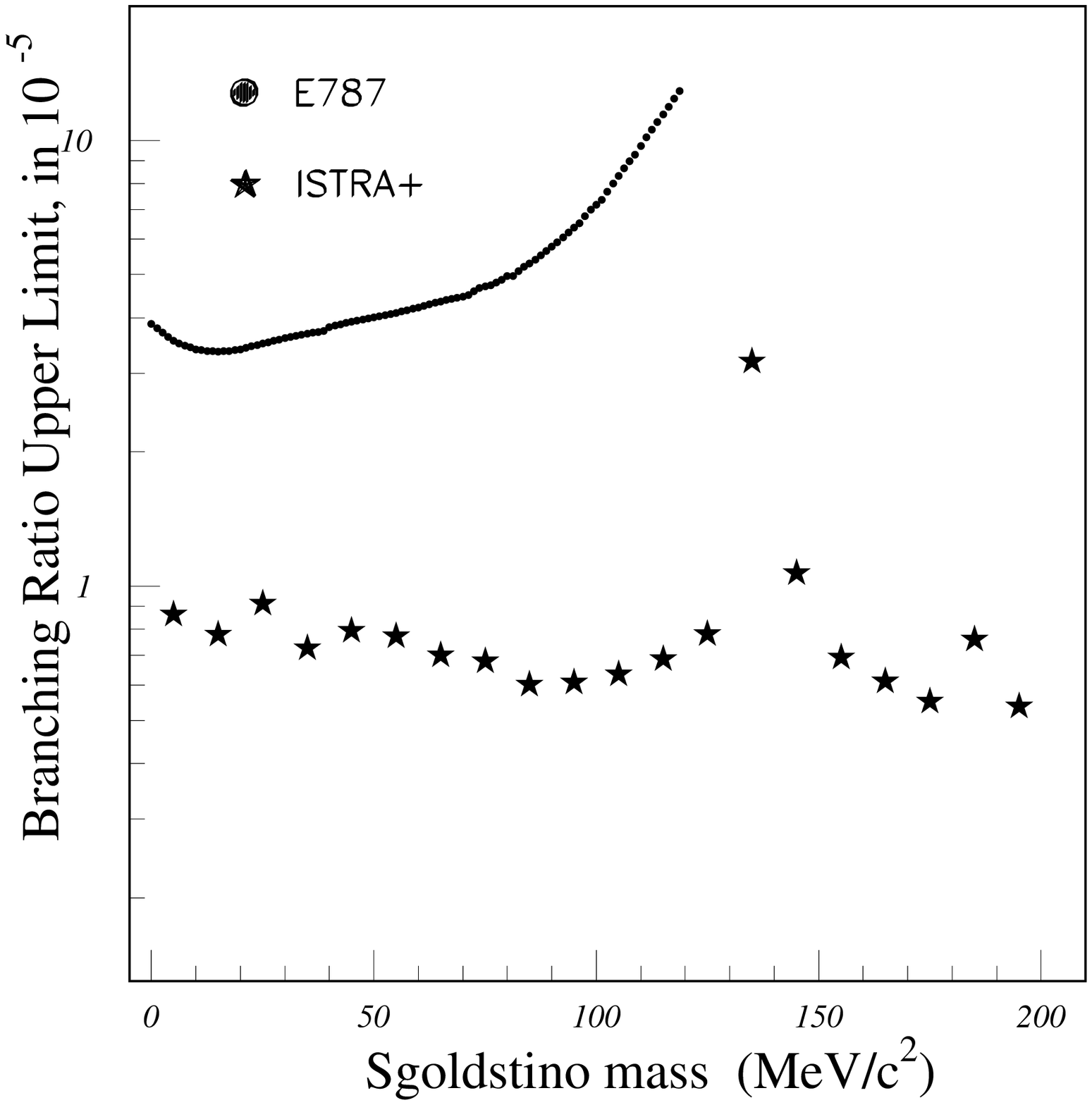,width=8cm}
&
\epsfig{file=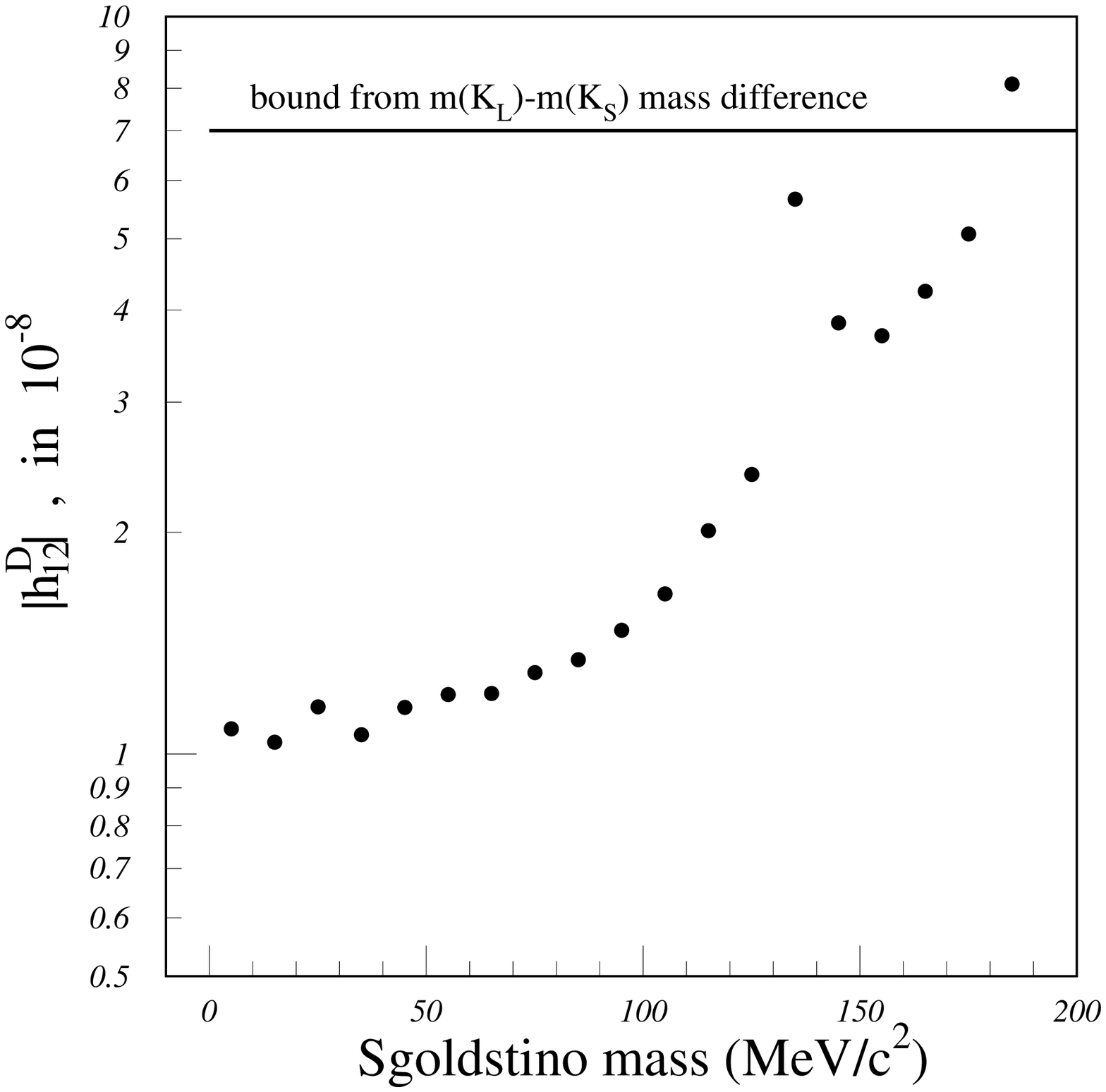,width=8cm}
\\
\end{tabular}
\caption{ The $90~\%$ CL upper limit for 
 the $Br(K^-\rightarrow \pi^-\pi^{\circ}P)$
 versus sgoldstino mass compared with the E787 upper limit(left),
 the $90~\%$ CL upper limit for the $|h^D_{12}|$   compared with 
 the theoretical 
 limit from $K_L - K_S$ mass difference(right).}
\end{figure} 

\section{Summary and conclusions}

 A search for the light pseudoscalar sgoldstino  in the 
 $K^{-} \rightarrow \pi^{-} \pi^{\circ} P$ decay is performed. 
It is assumed that
 sgoldstinos decay outside the detector.
 No signal is seen in the $m_{P}$ mass interval between 0 and 200~MeV/c$^2$.
 The   upper limits at 90$\%$~CL  are set to be
  $\sim 9.0~\cdot~10^{-6}$  
for the sgoldstino mass range
 from $0$ to $200$~MeV/c$^2$, excluding the interval near $m(\pi^{\circ})$,
 where the limit is $\sim 3.5~\cdot~10^{-5}$.
  These results  improve the confidence limits
 published by the E787 collaboration. Our limits improve also the
 theoretical constraints on $|h^D_{12}|$ from $K_L - K_S$ mass
 difference.  
  

The authors would like  to thank D.S.~Gorbunov, V.A.~Matveev and V.A.~Rubakov, 
for  numerous discussions.  

 The work 
 is  supported in part by the RFBR grants N03-02-16330 (IHEP group)
 and N03-02-16135 (INR group).


\end{document}